\documentclass[letter,traditabstract]{aa} 
\usepackage{graphicx}
\usepackage{txfonts}
\usepackage{rotating}
\usepackage{enumitem}
\usepackage{appendix}
\usepackage{longtable, lscape}
\begin{document}
\title{\emph{Herschel}\thanks{\emph{Herschel} is an ESA space observatory with science instruments provided by 
European-led Principal Investigator consortia and with important participation from NASA.}
FIR counterparts of selected Ly{$\alpha$} emitters at z$\sim$2.2}
\subtitle{Fast evolution since z$\sim$3 or missed obscured AGNs?}

\author{
\'A. Bongiovanni\inst{1,2}
\and
I. Oteo\inst{1,2}
\and
J. Cepa\inst{2,1}
\and
A.~M. P\'erez Garc\'{\i}a\inst{1,2}
\and
M. S\'anchez-Portal\inst{3}
\and
A. Ederoclite\inst{1,2}
\and
J.~A.~L. Aguerri\inst{1,2}
\and
E.~J. Alfaro\inst{4}
\and
B. Altieri\inst{3}
\and
P. Andreani\inst{14}
\and
M.~T. Aparicio-Villegas\inst{4}
\and
H. Aussel\inst{15}
\and
N. Ben\'{\i}tez\inst{4}
\and
S. Berta\inst{13}
\and
T. Broadhurst\inst{12}
\and
J. Cabrera-Ca\~no\inst{5}
\and
F.~J. Castander\inst{6}
\and
A. Cava\inst{1,2}
\and
M. Cervi\~no\inst{4}
\and
H. Chulani\inst{1}
\and
A. Cimatti\inst{16}
\and
D. Crist\'obal-Hornillos\inst{4,9}
\and
E. Daddi\inst{15}
\and
H. Dominguez\inst{17}
\and
D. Elbaz\inst{15}
\and
A. Fern\'andez-Soto\inst{10}
\and
N. F{\"o}rster Schreiber\inst{13}
\and
R. Genzel\inst{13}
\and
M.~F. G\'omez\inst{1}
\and
R.~M. Gonz\'alez Delgado\inst{4}
\and
A. Grazian\inst{17}
\and
C. Gruppioni\inst{18}
\and
J.~M. Herreros\inst{1}
\and
S. Iglesias Groth\inst{1,2}
\and
L. Infante\inst{11}
\and
D. Lutz\inst{13}
\and
B. Magnelli\inst{13}
\and
G. Magdis\inst{15}
\and
R. Maiolino\inst{17}
\and
I. M\'arquez\inst{4}
\and
V.~J. Mart\'{\i}nez\inst{7}
\and
J. Masegosa\inst{4}
\and
M. Moles\inst{4,9}
\and
A. Molino\inst{4}
\and
R. Nordon\inst{13}
\and
A. del Olmo\inst{4}
\and
J. Perea\inst{4}
\and
A. Poglitsch\inst{13}
\and
P. Popesso\inst{13}
\and
F. Pozzi\inst{18}
\and
F. Prada\inst{4}
\and
J.~M. Quintana\inst{4}
\and
L. Riguccini\inst{15}
\and
G. Rodighiero\inst{19}
\and
A. Saintonge\inst{13}
\and
S.~F. S\'anchez\inst{8,9}
\and
P. Santini\inst{17}
\and
L. Shao\inst{13}
\and
E. Sturm\inst{13}
\and
L. Tacconi\inst{13}
\and
I. Valtchanov\inst{3}
}
\institute{\centering \vskip -10pt \small \it (See online Appendix \ref{sect:affiliations} for author affiliations) }

\date{Received 2 April 2010 / Accepted 3 August 2010}

\offprints{\'Angel Bongiovanni, \email{bongio@iac.es}}
 
  \abstract{
   Ly{$\alpha$} emitters (LAEs) are seen everywhere in the redshift domain from local
   to z$\sim$7. Far-infrared (FIR) counterparts of LAEs at different epochs could provide
   direct clues on dust content, extinction, and spectral energy distribution (SED) for 
   these galaxies. We search for FIR counterparts of 
   LAEs that are optically detected in the GOODS-North field at redshift z$\sim$2.2 using data from 
   the \emph{Herschel} Space Telescope with the Photodetector Array Camera and Spectrometer (PACS). 
   The LAE candidates were isolated via color-magnitude diagram using the medium-band photometry 
   from the ALHAMBRA Survey, ancillary data on GOODS-North, and stellar population models. According 
   to the fitting of these spectral synthesis models and FIR/optical diagnostics, most of them 
   seem to be obscured galaxies whose spectra are AGN-dominated. From the analysis of the optical data, 
   we have observed a fraction of AGN or composite over source total number of $\sim$0.75 in the LAE population at z$\sim$2.2, 
   which is marginally consistent with the fraction previously observed at z=2.25 and even at low redshift (0.2$<$z$<$0.45), but 
   significantly different from the one observed at redshift $\sim$3, which could be compatible either with a 
   scenario of rapid change in the AGN fraction between the epochs involved or with a non detection of 
   obscured AGN in other $z=2-3$ LAE samples due to lack of deep FIR observations. We found three robust 
   FIR (PACS) counterparts at z$\sim$2.2 in GOODS-North. This demonstrates the possibility of finding 
   dust emission in LAEs even at higher redshifts.}

   \keywords{Infrared: galaxies -- Galaxies: evolution -- high redshift}

   \maketitle
%

\section{Introduction}\label{intro}

Ly$\alpha$ emitters (LAEs) are found within the more distant baryonic structures so
far detected in the universe. Like most high-redshift objects, they are classified
according to the detection technique, a procedure that has generated a wide
collection of acronyms (EROs, LBGs, SMG, etc) that are an indication of our lack
of knowledge of the galaxy evolution processes at high redshift. The LAEs can be
found at almost any redshift from local ({\"O}stlin et al. \cite{ostlin09},
Deharveng at al. \cite{deharveng08}) through z$\sim$7 (Bouwens et al. \cite{bouwens09},
Iye et al. \cite{iye06}) and beyond ({\it e.g.} Sobral et al. \cite{sobral09}; 
Bouwens et al. \cite{bouwens10}), 
although LAEs at low redshift (Finkelstein et al. \cite{finkelstein09a,finkelstein09d}) show quite 
different properties from those at z$>$2. 

On the high redshift side, LAEs are natural indicators of the
reionization of the universe, although the present evidence goes from the early
reionization models, which claim that reionization is nearly complete at z$\sim$8,
through the late reionization ones, to end around redshift 6.6 
(Choudhury \& Ferrara \cite{choudhury06}),
supported by a number density of LAEs that seem to decrease beyond z$\sim$6
(Kobayashi et al. \cite{kobayashi07}). On the low redshift extreme, at z$\sim$2-3, the
relatively scarce number of LAEs detected could be consistent with them being
the progenitors of present day L* galaxies. Gawiser et al. (\cite{gawiser07})
studied a sample of 162 LAEs at z$\sim$3.1 and found neither evidence 
of dust obscuration nor a substantial AGN component (of $\sim$1\%) in the host galaxies, which determines
that their sample is essentially composed of young, low stellar mass sources without
possible far-infrared (FIR) counterparts. These results agree with those obtained by Nilsson et al. (\cite{nilsson07}), who 
studied a stacked sample of LAEs at z=3.15. Nevertheless, there is recent evidence about 
LAEs with different dust contents. For instance, Pirzkal et al. (\cite{pirzkal07}) found dusty but
young LAEs at z$\sim$5. Even more recently, in a study of 170 robust LAE 
candidates at z=2.25, Nilsson et al. (\cite{nilsson09}) found a trend of apparent evolution in the LAE
properties with respect to their previous work: they detect a significant AGN contribution 
and red spectral-energy distributions, which imply more massive, dustier, and older galaxies 
than their {\it relatives} at z$\gtrsim$3. This result is stressed in Nilsson \& M$\o$ller (\cite{nilssonmoller09}),
who found that a non-zero fraction of LAEs at z$<$3 are ULIRGs.

Therefore, for a meaningful fraction of LAEs at z=2-3 it
is possible to obtain direct evidence of dust re-emission in FIR produced by the absorption 
of UV and optical photons from star-forming regions or nuclear activity. Additionally, the Ly$\alpha$
photons are resonantly scattered by neutral hydrogen in the galactic ISM, increasing the
probability that they are totally screened. Because of this, Hayes et al. (\cite{hayes10}) have recently 
confirmed that a huge fraction (almost a 90\%) of star-forming galaxies emit insufficient Ly$\alpha$ 
photons to be detected by narrow-band surveys. 

Even at higher redshift there is evidence of probably dusty LAEs. 
Finkelstein et al. (\cite{finkelstein09b}) performed an analysis of the expected detection 
of dust emission for high-z, narrow-band selected LAEs in GOODS {\it Chandra} Deep Field-South 
(CDF-S). Dust in a fraction of $\simeq$0.4 of LAEs with redshift between 4.1 and 5.8 could be detectable 
in rest-frame wavelengths of 60 and 100 $\mu\textrm{m}$. This result is reinforced in Finkelstein 
et al. (\cite{finkelstein09c}) from a standard SED fitting of LAEs at z$\sim$4.5 in the same field.
Additionally, they propose that the bimodality observed in the age distribution of LAE stellar 
populations may be owing to dust. With more conservative results, but using a previously developed 
Ly$\alpha$/continuum production/transmission model, Dayal et al. (\cite{dayal10}) claim to have found an 
efficient strategy to look for dust emission from LAEs using the new developments of ALMA: they 
suggest that the Ly$\alpha$ and submillimeter emissions (preferentially in the 850 $\mu\textrm{m}$ band) 
for galaxies at z=(5.7, 6.6; FIR in the LAE rest-frame) are correlated and, therefore, a
fraction of high-z LAEs (but significantly smaller than the predicted by Finkelstein et al.
\cite{finkelstein09c}) could be observed in the submillimeter regime. This could be supported by the spatial
correlation claimed between SMGs and LAEs, where both populations act as high density tracers
(Tamura et al. \cite{tamura09}). Thus, AGN fraction, dust contents, and correlation with FIR data
would help to provide clues for the evolution of LAEs.

From this brief review, it is clear that the high-z LAE properties are poorly known and on other hand, 
finding possible counterparts of these objects in the FIR and submillimeter regimes -as we try to demonstrate in this 
paper- can help to constrain the nature of this apparent duality of LAEs.
We present the first results of a multiwavelength analysis of LAE candidates
with observations performed with the ESA \emph{Herschel} Space Observatory (Pilbratt et al.
\cite{pilbratt10}) and the PACS instrument (Poglitsch et al. \cite{poglitsch10}) in the framework
of the PACS Evolutionary Probe (PEP, PI D. Lutz). The PEP is the \emph{Herschel} Guaranteed Time Key-Project
designed to obtain the best profit from \emph{Herschel} instrumentation to study the FIR galaxy
population. In our case, only very strong and dusty LAEs could be detected with PACS.
Finally, a relative fraction of LAEs hosting AGNs with respect to the overall population is estimated.

Throughout this paper a concordant cosmology with ${\rm H}_0=70\ {\rm km}\ {\rm s}^{-1}\ {\rm Mpc}^{-3}$ is assumed.
Unless otherwise specified, all magnitudes are given in the AB system.

\section{Sample selection and ancillary data}

\subsection{Optical identification of candidate LAEs}

We searched for FIR counterparts of LAEs at ${\rm z} \sim 2.2$ in the northeastern
half of the GOODS-North field ($\sim$70 sq-arcmin) by using selected filters of the ALHAMBRA system
as a part of a more extended study. 
The Advanced Large Homogeneous Area Medium-Band Redshift Astronomical (ALHAMBRA) survey is
aimed at providing a tomography of the evolution of the contents of the universe over
most of their cosmic history (see Moles et al. \cite{moles08}
for a more detailed description of the survey and its scientific goals).
This novel approach employs 20 contiguous, equal-width,
FWHM$\sim$320 \AA\ top-hat filters covering from 3500 to 9700 \AA\, plus the
Johnson-standard JHKs near-infrared (NIR) bands, to observe a total area of 3.5 ${\rm deg}^2$
on the sky (a description of the ALHAMBRA photometric system is given in Aparicio Villegas et 
al. \cite{apariciov10}). 
The observations were carried out with the Calar Alto 3.5 m telescope using the wide-field cameras in the
optical, Large Area Imager for Calar Alto (LAICA), and in the NIR, Omega-2000. The magnitude
limits achieved by ALHAMBRA are AB = 25.5 mag (for an unresolved object, the signal-to-noise ratio S/N=5)
in the optical filters from the blue to 8300 \AA, and from AB = 24.7 to 23.4 for the redder ones.
The limits in the NIR are in the Vega system J$\sim$22 mag, H$\sim$21 mag, and Ks$\sim$20 mag.

\begin{figure}
\centering
\resizebox{\hsize}{!}{\includegraphics{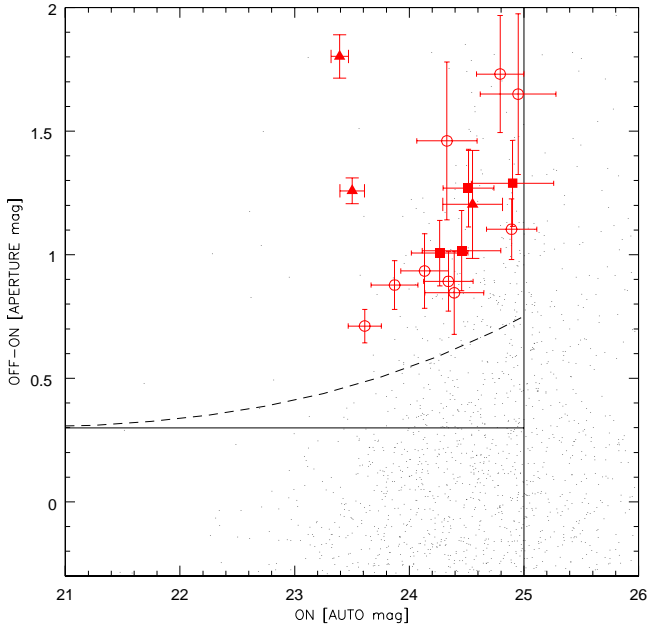}}
\caption{Diagnostic diagram of the whole catalog and candidates (see online
Appendix \ref{sect:z2LyaCandidatesSelection} for a definition of ON and OFF bands). Continuous lines represent the basic 
cut in the color OFF-ON and magnitude in the ON-band. Dashed curve indicates the 3-$\sigma$ color selection threshold, above
which the LAE candidates are located. The objects with error bars represent the final sample of LAE candidates:
the filled triangles represent LAEs with MIPS (24 $\mu\textrm{m}$ band) and PACS (100/160 $\mu\textrm{m}$ bands) 
counterparts, the filled squares are LAEs with FIR detection in MIPS data only, and the open symbols are 
the LAEs without FIR counterparts at the sensitivity limit of PEP SDP data.
}
\label{CMD}
\end{figure}

The searching procedure adopted is similar to the well-known narrow-band techniques used to find 
high-redshift galaxies (Cowie \& Hu \cite{cowie98},
Gronwall et al. \cite{gronwall07}, Ouchi et al. \cite{ouchi08}, Shioya et al. \cite{shioya09},
Murayama et al. \cite{murayama07}), but instead of a
combination of narrow and broad filters to isolate the line and define the continuum,
we used selected ALHAMBRA intermediate bandpass filters for both purposes. Details of the methodology 
are given in the online Appendix \ref{sect:z2LyaCandidatesSelection}.
Figure \ref{CMD} shows a detail of the medium-band color-magnitude distribution for our catalog of 2532 spurious-free
sources, detected in the northeast fraction of the GOODS-North field.
After applying the color- and limiting 
magnitude selection criteria defined in Appendix \ref{sect:z2LyaCandidatesSelection}, with an additional 3-$\sigma$ 
color-magnitude restriction (dashed line in Fig. \ref{CMD}), we found 134 raw LAE candidates. This gives a mean number 
density of $\sim2\times10 ^{-3}\ \textrm{Mpc}^{-3}$.
However, without spectroscopic information for a statistically significant sample of our raw LAE candidates,
these sources were fitted with galaxy templates BC03 (Bruzual \& Charlot \cite{bruzual03}), using the procedure described in 
Appendix \ref{sect:z2LyaCandidatesSelection} to discard 
the continuum-only objects that show a color excess. Individual fittings also allow us to obtain reliable 
photometric redshifts and preliminary spectral classifications for the raw LAE candidates. For this purpose, we 
adopted the photometry from Capak et al. (\cite{capak04}) in the optical (UBVRIz') instead of the ALHAMBRA one. The 
former data are about 1.1 to 1.5 mag (AB) deeper in U, B, V, and R bands than the latter. Not so for the NIR 
photometry, where ALHAMBRA data are given in the canonical bands. After applying this procedure, we combined 
the results with an analysis of detection reliability of the Ly$\alpha$ emission line and found 16 secure
candidates to LAEs, which are represented in the color-magnitude diagram of Fig. \ref{CMD} with their corresponding 
error bars. Optical pseudo-spectra from ALHAMBRA survey and relevant data of these LAEs, including the
estimated Ly$\alpha$ luminosity, are given in online Fig. \ref{z2LyaCandidatesSpectra} and Table \ref{z2LyaCandidatesTable},
respectively.

Additionally, each pseudo-spectrum is 
complemented with a 3$\times$3 sq-arcsec cutout in z-band from HST-ACS (i.e. close to the UV rest-frame 
of the source, avoiding possible clumpy features in the far-UV images), when available. With these data we estimated 
sizes and morphologies of the sources from the isophotal radius (2-$\sigma$ above background) and fitted one-component 
S\'ersic profiles using {\tt GALFIT} 
(Peng et al. \cite{peng02}). In all cases, this approach converged succesfully. The results of this 
analysis are also included in Table \ref{z2LyaCandidatesTable}. The LAEs at z$\sim$2.2 have a mean radius of $1.7\pm0.4$ kpc,
except for the objects ALH06146 and ALH07181, which exhibit S\'ersic profiles and residuals that suggest the
presence of bars and out-of-mean radii. Apart from these two objects, sources in the sample are essentially
compact and their S\'ersic indexes are consistent with bulge-like galaxies.

\subsection{Matching candidates with FIR data and modeling}

We took advantage of the availability of PACS data on the GOODS-North field to search for the counterparts of the 
final LAE sample in the FIR (100 and 160$\mu\textrm{m}$ PACS bands, with a sensitivity of 
$\sim$5.1 and $\sim$8.7 mJy at 5-$\sigma$, 
respectively) by using the PEP Science Demonstration Phase (SDP) catalog with MIPS-24 $\mu\textrm{m}$ based 
position priors (Berta et al. \cite{berta10}). We performed a match between the catalogs to find which candidates 
have a detected FIR source closer than $1.5 \,\textrm{arcsecs}$ (in the order of the pixel size at
100 $\mu\textrm{m}$), 
and thus study the resulting sample with the aid of spectral synthesis templates from Polletta et al. (\cite{polletta07}) 
for star-forming (SF) and AGN/Composite (AGN/C) objects, but adding the FIR fluxes to the photometric data set previously 
used. In this preliminary analysis we did not use mid-infrared (MIR) data from {\it Spitzer}-IRAC because not 
all the LAE candidates were detected in the raw images. Even so, the IRAC-bands fluxes using ALHAMBRA Ks-band catalog 
as priors, with the same aperture set, will be included in a forthcoming analysis of PEP data (Oteo et al. 2010, in prep.)

\begin{figure}
\centering
\resizebox{\hsize}{!}{\includegraphics{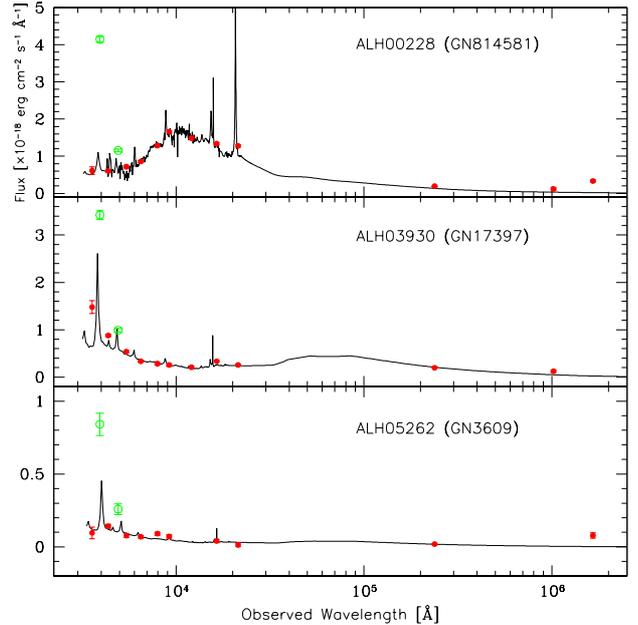}}
\caption{Best-model fitting of the final LAE candidates at z$\sim$2.2 with PACS counterparts. SWIRE templates of 
AGN \& composite sources from
Polletta et al. (\cite{polletta07}) fit the LAE UBVRIz' photometry from Capak et al. (\cite{capak04}) and the JHKs one 
from ALHAMBRA (Moles et al. \cite{moles08}). The red dots represent the photometric data fitted. The open green circles 
represent the aperture fluxes measured in the ON-band (A394M filter, used to account for the redshifted Ly$\alpha$ flux) 
and through the A491M filter (in which wavelength range the redshifted CIV$\lambda$1549 line would be contained). The
latter values were superposed on the plots, {\it i.e.}, they were not included in the fitting.
}
\label{fit_models}
\end{figure}

We found that seven out of 16 LAEs are detected in MIPS 24 $\mu\textrm{m}$ band, of which three were also detected at 
100 and/or 160$\mu\textrm{m}$ bands of PACS. The best-fit for one source of our sample of LAEs with MIPS-24 $\mu\textrm{m}$ 
counterparts corresponds to a SF-like template, and the remaining six sources were well fitted with AGN/Composite ones. The
LAEs with FIR counterparts in PACS data belong to the latter group, and in the following discussion we only
consider those galaxies, unless otherwise specified. The fluxes of the sources in FIR as well 
as the final spectral classification derived from the best-fitting templates are given in 
Table \ref{z2LyaCandidatesTable}. We also distinguished these LAEs in the color-magnitude distribution 
given in Fig. \ref{CMD}. Likewise, the best-fitted spectra for the LAEs with FIR counterparts in PACS data,
corresponding to the objects identified as ALH00228, ALH03930 (with spectroscopic redshift), and ALH05262, are 
represented in Fig. \ref{fit_models}. The best fitting obtained for the first object corresponds to an AGN-1/Composite 
spectrum, while the other two were fitted using AGN-2/Composite ones. As a consistency test, we compared the 
F$_{\nu}(100 \ \mu{\rm m})$/F$_{\nu}(24 \ \mu{\rm m})$ ratios for the objects ALH00228 and ALH03930
(11.51$\pm$3.46 and 11.57$\pm$2.31, respectively) with the AGN/SB diagnostic predictions of Mullaney 
et al. (\cite{mullaney10}) for PACS filters. These sources fall on the average region (at z=2.2) of AGN composed with
SBs.

Finally, we matched the final LAE catalog with the X-ray data of {\it Chandra}/GOODS-North from Alexander et al.
\cite{alexander03}. We found only one counterpart (object ALH08364) in this catalog, whose best-fitted template 
correspond to AGN/Composite.

\section{Results and discussion}

We found 134 raw candidates at a mean redshift of 2.2 in the northeastern half of GOODS-North field, using
color-magnitude diagnostics of selected medium-band data from ALHAMBRA survey. From this sample, we segregated
16 robust LAEs using spectral synthesis templates. 75\% of the final sample were well fitted with AGN/Composite
templates, whereas the remaining galaxies were star-forming. Likewise, almost half of the LAEs detected at z$\sim$2.2 
have MIPS 24 $\mu\textrm{m}$ counterparts ($\sim$7.5 $\mu\textrm{m}$ rest-frame), of which three sources have PACS 
100 and/or 160 $\mu\textrm{m}$ counterparts ($\sim$31 and/or 50 $\mu\textrm{m}$ rest-frame). This proves the possibility 
of finding dust emission evidence at high redshift, taking into account the sensitivity limits of the PEP-SDP data 
(100 $\mu\textrm{m}$ band: $\sim$5.1 mJy, 5$\sigma$). Moreover, using the the conversion from MIR to total
infrared luminosity of Chary \& Elbaz (\cite{charyelbaz01}, eq. 6), we found that the three brightest 
counterparts of our LAEs in the 24 $\mu\textrm{m}$ band are ULIRGs, which partially agree with the main result 
of Nilsson \& M$\o$ller (\cite{nilssonmoller09}) for LAEs at redshifts below z=3. In addition, positive finding of FIR 
counterparts of spectroscopically observed LAEs has been recently confirmed by our team, using the first data release 
of PEP Data on the GOODS-South field (Oteo et al. 2010, in prep.).

From the analysis of available images of HST-ACS in z-band (rest-frame UV), the LAEs at z$\sim$2.2 can be
described as mainly compact (2 out of 16 of barred morphology), with a typical isophotal radius of $\sim$1.7 kpc. This
finding is consistent with the dominant morphology in the LAE sample at z=3.1 analyzed by Bond et al. (\cite{bond09}), but
with sizes larger by a factor $\sim$2.

At Ly$\alpha$ luminosity and rest-frame EW limits of 1.95$\times$10$^{42}$ erg s$^{-1}$ and 35$\AA$, respectively, and 
under the cosmological assumptions stated in Sect. \ref{intro}, we compared our LAE sample of possible AGNs with those 
from recently published data at z=2.25 of Nilsson et al. (\cite{nilsson09}). To search for possible evolutionary effects 
on the AGN fraction, we also analyzed the LAE samples of Gronwall et al. (\cite{gronwall07}) and 
Ouchi et al. (\cite{ouchi08}), both at z=3.1. The AGN-to-total fraction for each sample is given 
in Table \ref{AGNfractionTable}. 
The position of each source of these samples in the previously defined EW-luminosity space is shown in Fig.
\ref{AGNfractionPlot}. From these data it is clear that the AGN fraction increased by a factor of 2.5 
(Nilsson et al. \cite{nilsson09}) to 15 (this work) between both epochs ($\Delta{\rm t}\sim$0.9 Gyr), assuming 
a mean AGN fraction at z=3.1 of 0.05. This result would suggest a rapid evolution scenario of LAEs classified as AGNs 
between z$\sim$2 and 3. Wolf et al. (\cite{wolf03}) found a peak in the comoving AGN space density at z$\sim$2, but 
our {\it prima facie} evidence points to an evolutionary behavior that qualitatively exceeds the one reported by 
them. Additionally, our AGN-to-total ratio in the LAE population at z$\sim$2.2 is marginally consistent with that 
observed at low redshift (of 0.435 at 0.2$<$z$<$0.45, Finkelstein et al. \cite{finkelstein09a}). 
Note that the AGNs in the samples used for comparison purposes have been classified as such 
because of the X-ray detection whereas in our case, the lack of detection of FIR counterparts in X-rays 
({\it Chandra}/GOODS-North, Alexander et al. \cite{alexander03}) suggests that these sources are obscured AGNs. 
Consistently, at z$\sim$2.2 the FIR emission between 30 to $\sim$110 $\mu\textrm{m}$ (rest-frame) could be attributed 
to the re-emission of the dust heated by either the active nucleus or starburst regions close to the nucleus 
(i.e. a {\it warm} dust component; P\'erez Garc\'{\i}a et al. \cite{perez98}). If only X-ray detected AGN were 
considered, the fraction of active nuclei within our sample would be reduced to $\sim$\,0.16, i.e. compatible 
with no evolution from the z\,$\simeq$\,3.1 samples.

As an alternative to the presented evolutionary scenario, one could suggest the presence of some kind of selection effect.
As shown in Fig. \ref{AGNfractionPlot}, the sources of our whole LAE sample with $\log$ L(Ly$\alpha$)$>$42.5 
have rest-frame EWs a factor of 2-3 higher than the corresponding objects in the sample of Nilsson et al. (\cite{nilsson09})
at the same redshift. A possible explanation for this effect is that our ON-filter could favor the selection of 
broad emission-line objects, but this does not clarify the absence of AGNs with rest-frame EWs above $\sim$150$\AA$
in the latter. 

This assertion is strengthened by the evidence of the CIV$\lambda$1549 redshifted emission line in the ALHAMBRA optical 
pseudo-spectra of these sources (observed through the filter A491M), apart from the photometric signature of the Ly$\alpha$ 
emission, as shown in Fig. \ref{fit_models}. Moreover, the R-Ks colors and F$_{\nu}(24 \ \mu{\rm m})$/F$_{\nu}$(R) ratios (see 
Table \ref{z2LyaCandidatesTable}) of the FIR counterparts of our LAE sample give preliminary evidence about
the degree of obscuration: many of our sources could host moderately to highly obscured AGNs, according to the criteria
extensively discussed in Fiore et al. (\cite{fiore08}), if the approximation K-band$\approx$Ks is allowed. Thus,
returning to the dicothomy previously raised and apart from the possible evolution of the AGN fraction between
redshifts 2 and 3, one might consider that some sources at z=2-3, classified as LAEs by different authors, are also 
obscured AGNs.
 
\begin{acknowledgements}

This work was supported by the Spanish Plan Nacional de Astrononom\'{\i}a y Astrof\'{\i}sica 
under grants AYA2008-06311-C02-01 and AYA2006-14056. Based on observations collected at the German-Spanish 
Astronomical Center, Calar Alto, jointly operated by the Max-Planck-Institut für Astronomie, 
Heidelberg and the Instituto de Astrof\'{\i}sica de Andaluc\'{\i}a (CSIC). PACS has been developed 
by a consortium of institutes led by MPE (Germany)
and including UVIE (Austria); KUL, CSL, IMEC (Belgium); CEA, OAMP
(France); MPIA (Germany); IFSI, OAP/AOT, OAA/CAISMI, LENS, SISSA
(Italy); IAC (Spain). This development has been supported by the funding agencies
BMVIT (Austria), ESA-PRODEX (Belgium), CEA/CNES (France), DLR
(Germany), ASI (Italy) and CICYT/MICINN (Spain). We thanks the anonymous referee for valuable comments, 
which have contributed significatively to the manuscript improvement.

\end{acknowledgements}

{}

\Online

\appendix

\section{Author's affiliations}\label{sect:affiliations}

\begin{enumerate}[label=$^{\arabic{*}}$]

\item Instituto de Astrof{\'i}sica de Canarias (IAC), E-38200 La Laguna, Tenerife, Spain
\item Departamento de Astrof{\'i}sica, Universidad de La Laguna (ULL), E-38205 La Laguna, Tenerife, Spain
\item Herschel Science Centre (ESAC). Villafranca del Castillo, Spain
\item Instituto de Astrof{\'i}sica de Andaluc{\'i}a (CSIC), Granada, Spain
\item Departamento de F{\'i}sica At\'omica, Molecular y Nuclear, Facultad de F{\'i}sica, Universidad de Sevilla, Spain
\item Institut de Ci{\`e}ncies de l'Espai (CSIC), Barcelona, Spain
\item Departament d'Astronomía i Astrof{\'i}sica, Universitat de Val{\`e}ncia, Val{\`e}ncia, Spain
\item Centro Astron\'omico Hispano-Alem\'an, Almer{\'i}a, Spain
\item Centro de Estudios de F{\'i}sica del Cosmos de Arag\'on, CEFCA, E-44001 Teruel, Spain
\item Instituto de F{\'i}sica de Cantabria (CSIC-UC), E-39005 Santander, Spain
\item Departamento de Astronom{\'i}a, Pontificia Universidad Cat\'olica, Santiago, Chile
\item School of Physics and Astronomy, Tel Aviv University, Israel
\item Max-Planck-Institut f\"{u}r Extraterrestrische Physik (MPE), Postfach 1312, 85741 Garching, Germany
\item ESO, Karl-Schwarzschild-Str. 2, D-85748 Garching, Germany
\item Laboratoire AIM, CEA/DSM-CNRS-Universit{\'e} Paris Diderot, IRFU/Service d'Astrophysique, B\^at.709, CEA-Saclay, 91191 Gif-sur-Yvette Cedex, France
\item Dipartimento di Astronomia, Universit{\`a} di Bologna, Via Ranzani 1, 40127 Bologna, Italy
\item INAF - Osservatorio Astronomico di Roma, via di Frascati 33, 00040 Monte Porzio Catone, Italy
\item INAF - Osservatorio Astronomico di Bologna, via Ranzani 1, I-40127 Bologna, Italy
\item Dipartimento di Astronomia, Universit{\`a} di Padova, Vicolo dell'Osservatorio 3, 35122 Padova, Italy

\end{enumerate}

\section{Photometric selection of z$\sim$2.2 LAE candidates from ALHAMBRA survey in GOODS-North}\label{sect:z2LyaCandidatesSelection}

\begin{figure}[!h]
\centering
\resizebox{\hsize}{!}{\includegraphics{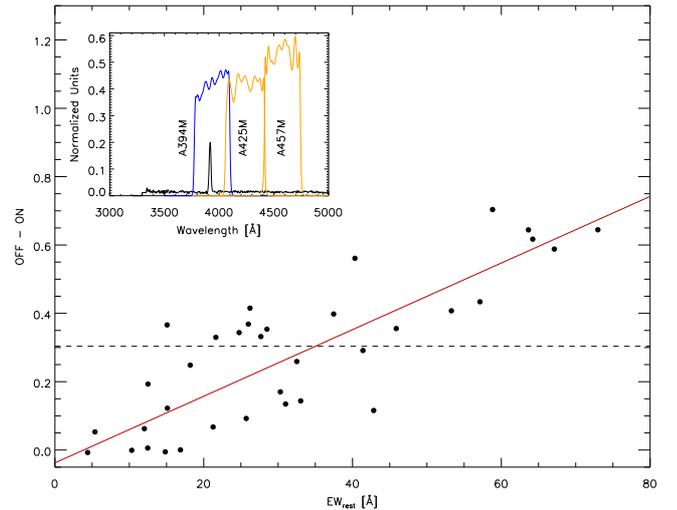}}
\caption{ALHAMBRA expected OFF-ON color of the VIMOS LAEs spectra at ${\rm z} \sim 3$, shifted to z=2, as a
function of their equivalent widths. The relation between both variables allows us to have a color
selection criterion. The dashed line represents the basic color threshold which corresponds to a
${\rm EW}_{\rm{rest}}= 35 \, \rm{\AA}$. Inset: Bandpasses of the filters used to find LAEs at $2.1<{\rm z}<2.37$.
The curves represent the ON (A394M) and OFF (A425M+A457M) filters transmissions, multiplied by the detector
quantum efficiency and the atmospheric transmission at airmass=1. A typical spectrum of a LAE at ${\rm z}=2.2$ is
overlaid to show the basis of the photometric selection criterium.}
\label{filter}
\end{figure}

The method used for finding LAEs is based on a color-magnitude diagnostic diagram.
The filter which samples the emission line is the ON filter and those used to define the continuum 
constitute the OFF filters set. The choice of the ON filter determines the range where the redshifts of the candidates 
will be. In this way, we seleted the LAE candidates at $2.10<{\rm z}<2.37$ by using the filters
A394M as ON filter (owing to the official naming of ALHAMBRA filters, the number between letters
gives the rounded central wavelength in nanometers; the last letter is the acronym of ``Medium'' bandpass),
and a sum of A425M and A457M as OFF filters, in order to increase the S/N in the continuum.
Figure \ref{filter} (inset) shows the
transmission curve of the chosen filters for LAE selection. Assuming an effective filter width equal
to its FWHM, and taking into account the sky area surveyed, as well as the emission line of interest, we
have explored in this way a comoving volume of about $6.63\times10^{4}\ \textrm{Mpc}^3$.

Once the ON and OFF filters were chosen, and according with the ALHAMBRA survey sensitivity, we adopt a
limiting magnitude of 25.0 in the ON-band. Then, we elaborated a color selection criterion to find LAEs
as efficiently as possible. We used the data from the GOODS/VIMOS Spectroscopy DR 2.0.1
(Popesso et al. \cite{popesso09}) to simulate the behavior of the SED from a typical LAE against the selected
ALHAMBRA filters. From the whole spectroscopic catalog we selected the
objects that show a line which corresponds to a Ly$\alpha$ emission. Then the spectra were de-redshifted
so that the central wavelength of the lines were
within the transmission range of the ON filter in each case, as shown in the example of Fig. \ref{filter}, and
their rest frame equivalent widths were calculated. After this we were
able to compute the OFF-ON color by convolution of the filter profiles with
each spectrum. In Fig. \ref{filter} we represent this
color against the rest frame equivalent width of the lines, measured with {\tt splot} in {\tt IRAF}.
As can be seen, there exists a relation between both variables, which allows us to define a color selection
criterion. Our LAE candidates were selected to have rest-frame equivalent widths
${\rm EW}_{\rm{rest}}\gtrsim 35\,\rm{\AA}$. According to Fig. \ref{filter}, this value corresponds to a
threshold color of approximately 0.3 for both filter pairs. But taking into account the photometric
errors, our adopted color selection criterion is
$m_{OFF}-m_{ON}-0.3\geq 3\sqrt{\sigma_{OFF}^2+\sigma_{ON}^2}$. This translates the minimum rest-frame equivalent
width to ${\rm EW}_{\rm{rest}}\gtrsim 35\,\rm{\AA}$ at the limiting magnitude in the ON-band, corresponding
to Ly$\alpha$ luminosities of $\log$ L=42.29 erg s$^{-1}$ at z=2.2.

All photometric measures were performed with {\tt SExtractor} (Bertin \& Arnouts \cite{bertin96}) on the full 
processed and stacked images of the field ALHAMBRA-5, pointing 1. To calculate the
color we used three arcsec aperture magnitudes and MAG\_AUTO ones to plot the magnitudes of the objects. Sources with
{\tt SExtractor} nonzero flags were discarded.

\begin{figure}[!h]
\centering
\resizebox{\hsize}{!}{\includegraphics{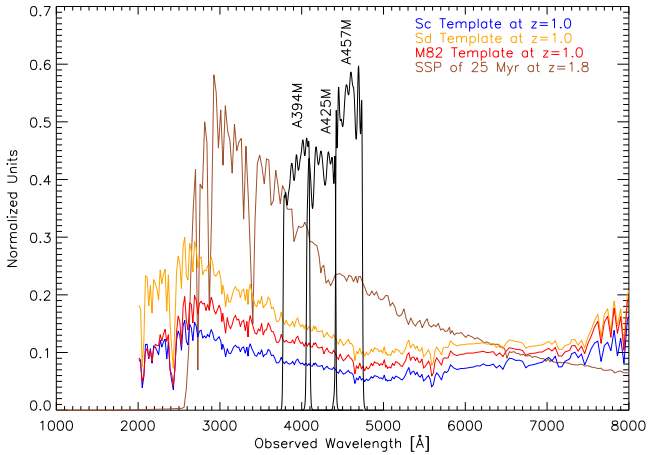}}
\caption{Spectra of possible contaminants. The transmission curves of the filter set used to select LAEs and
spectra of possible contaminants at different redshifts, built from BC03 and SWIRE
templates, are shown. The main contaminants are those whose UV continuum slope is sampled by the
filters, resulting in the appearance of false-positive candidates. For the sake of clarity, the spectra and the
transmission of the filters have been scaled.}
\label{contaminants}
\end{figure}

In order to study the nature of the possible continuum contaminants we carried out simulations with galaxy templates
of BC03 (Bruzual \& Charlot \cite{bruzual03})
and SWIRE (Polletta et al. \cite{polletta07}). We took all the templates and redshifted them from z=0 to 3 in bins of
$\Delta$z=0.05. We calculated their expected color and checked whether they satisfy the color
selection criterion. As a result, we found that the contaminants appear to be mainly starburst galaxies
at $1.0 \lesssim z \lesssim 2.0$, whose UV rest-frame drop is sampled with the filters. As an example, 
Fig. \ref{contaminants} shows a set of spectra of different starburst templates that satisfy the color 
criterium, although they are false positive candidates. More details
of this study are included in a forthcoming paper (Oteo et al. 2010, in prep.).

\section{Data of z$\sim$2.2 LAE candidates in GOODS-North}\label{sect:z2LyaCandidatesData}

\begin{figure*}[!ht]
\centering
\resizebox{\hsize}{!}{\includegraphics{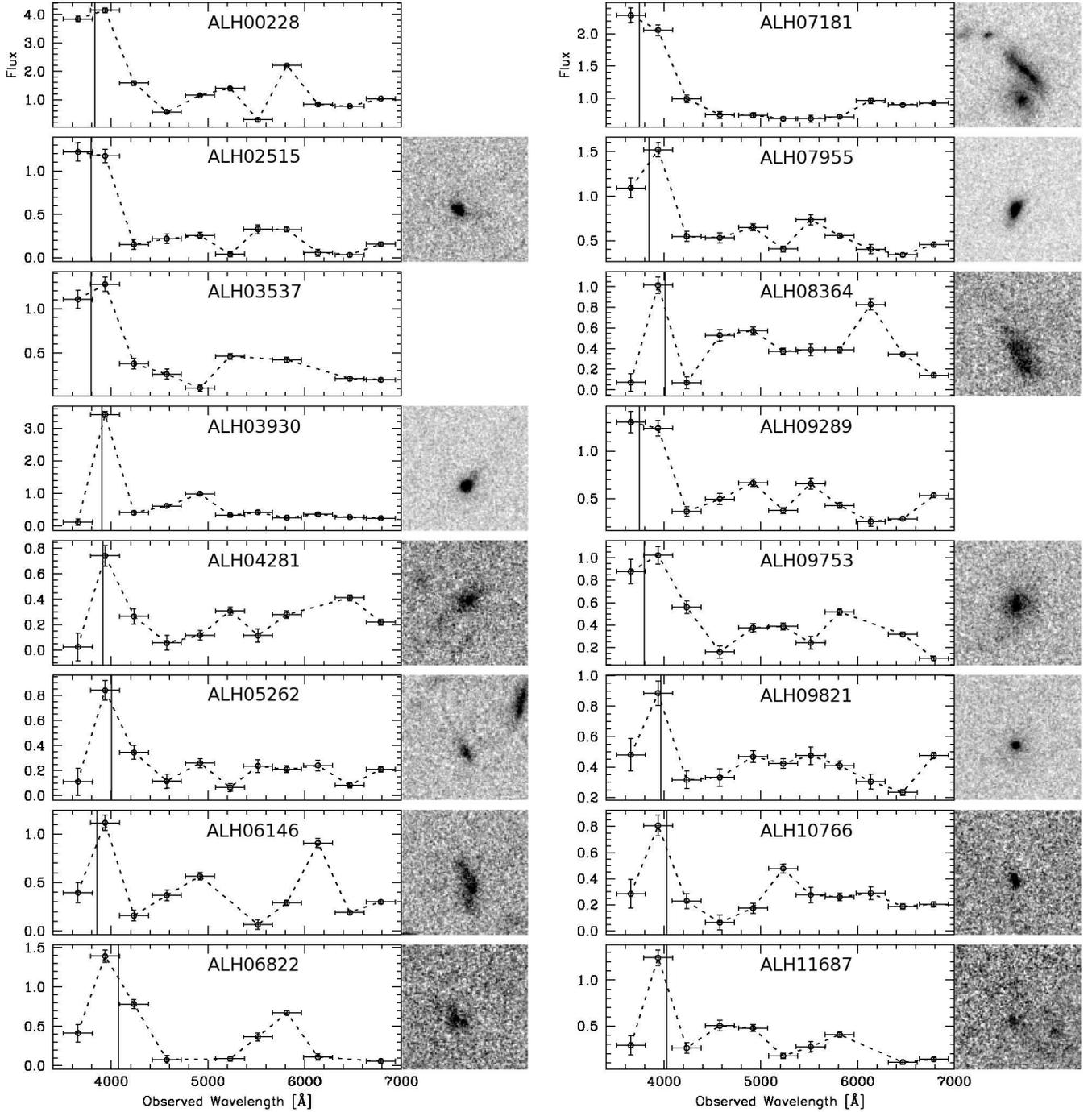}}
\caption{Optical pseudo-spectra of z$\sim$2.2 LAE candidates from ALHAMBRA survey. ACS cutouts (3$\times$3 sq-arcsec)
of the objects, when available, are shown in the right side of each pseudo-spectrum. Vertical line inside each panel
represent the position of Ly$\alpha$ emission line center at the corresponding photometric or spectroscopic redshift
given in Table \ref{z2LyaCandidatesTable}.
Fluxes are in 10$^{-18}$ erg cm$^{-2}$ s$^{-1}$ \AA$^{-1}$ units.
The horizontal bars represent the effective widths of ALHAMBRA filters. Assumming a concordant cosmology,
with ${\rm H}_0=70\ {\rm km}\ {\rm s}^{-1}\ {\rm Mpc}^{-3}$, the mean scale of the images is $\sim$8.2 kpc
arcsec$^{-1}$.}
\label{z2LyaCandidatesSpectra}
\end{figure*}

\begin{landscape}
\begin{table}[!ht]
\tiny
\caption{Combined data list of z$\sim$2.2 LAE candidates in the north-east half of the GOODS-North field.}
\label{z2LyaCandidatesTable}
\setlength{\tabcolsep}{0.06cm}
\begin{tabular*}{23.2cm}{ccccccccccccccccc}
\hline\hline
\tablefootmark{1}ID&RA&Dec.&\tablefootmark{2}z$_{\rm phot}$&\tablefootmark{3}z$_{\rm spec}$&\tablefootmark{4}EW(Ly$\alpha$)$_{\rm rest}$&\tablefootmark{5}OFF-ON&\tablefootmark{6}ON&\tablefootmark{7}$\log$ L(Ly$\alpha$)&\tablefootmark{8}F$_{24\ \mu{\rm m}}$&\tablefootmark{9}F$_{100\ \mu{\rm m}}$&\tablefootmark{10}F$_{160\ \mu{\rm m}}$&\tablefootmark{11}R$_{\rm isoph.}$&\tablefootmark{12}n&\tablefootmark{13}R-Ks&\tablefootmark{14}F$_{\nu}(24 \ \mu{\rm m}$)/F$_{\nu}$(R)&\tablefootmark{15}Sp.~T.\\
\hline
ALH00228&12:37:36.1&+62:10:58&2.15&-&159.50$\pm$14.36&1.259$\pm$0.052&23.502$\pm$0.108&43.02&0.192$\pm$0.005&0.120$\pm$0.031&0.336$\pm$0.029&-&-&3.33$\pm$0.06&77.38$\pm$2.45&AGN/C\\
ALH02515&12:37:34.3&+62:13:21&2.12&-&274.96$\pm$23.21&1.731$\pm$0.237&24.793$\pm$0.206&42.79&-&-&-&1.92&3.8&4.08$\pm$0.07&-&AGN/C\\
ALH03537&12:38:04.4&+62:14:26&2.12&-&161.72$\pm$14.52&1.270$\pm$0.157&24.517$\pm$0.222&42.61&0.013$\pm$0.003&-&-&-&-&4.96$\pm$0.06&15.48$\pm$3.94&AGN/C\\
ALH03930&12:37:04.3&+62:14:47&2.13&2.21&297.24$\pm$24.94&1.802$\pm$0.088&23.393$\pm$0.076&43.40&0.198$\pm$0.003&0.124$\pm$0.023&-&1.79&1.9&5.08$\pm$0.06&517.75$\pm$45.69&AGN/C\\
ALH04281&12:37:52.6&+62:15:07&2.22&-&202.84$\pm$17.66&1.461$\pm$0.319&24.328$\pm$0.263&42.86&-&-&-&1.70&1.9&3.91$\pm$0.07&-&AGN/C\\
ALH05262&12:37:48.3&+62:16:10&2.29&-&149.09$\pm$13.57&1.204$\pm$0.219&24.552$\pm$0.261&42.63&0.019$\pm$0.003&-&0.079$\pm$0.018&1.32&1.4&3.23$\pm$0.14&174.82$\pm$39.80&AGN/C\\
ALH06146&12:37:14.9&+62:17:00&2.17&2.13&165.52$\pm$14.81&1.289$\pm$0.174&24.901$\pm$0.359&42.49&0.014$\pm$0.003&-&-&0.64&0.8&4.32$\pm$0.07&48.95$\pm$12.63&AGN/C\\
ALH06822&12:37:43.2&+62:17:46&2.35&-&131.25$\pm$12.23&1.103$\pm$0.122&24.893$\pm$0.219&42.44&-&-&-&1.04&1.2&2.24$\pm$0.16&-&SF\\
ALH07181&12:37:30.7&+62:18:05&2.08&-&75.64$\pm$8.17&0.712$\pm$0.068&23.612$\pm$0.144&42.46&-&-&-&4.98&0.7&4.87$\pm$0.03&-&AGN/C\\
ALH07955&12:37:26.1&+62:18:57&2.16&-&96.76$\pm$9.68&0.877$\pm$0.099&23.872$\pm$0.205&42.57&-&-&-&2.74&1.2&4.26$\pm$0.05&-&SF\\
ALH08364&12:37:02.1&+62:19:25&2.30&-&117.16$\pm$11.18&1.017$\pm$0.163&24.456$\pm$0.343&42.53&0.037$\pm$0.003&-&-&1.78&0.7&3.97$\pm$0.06&72.09$\pm$8.56&AGN/C\\
ALH09289&12:37:41.6&+62:20:28&2.08&-&98.82$\pm$9.83&0.892$\pm$0.120&24.341$\pm$0.216&42.36&-&-&-&-&-&-&-&AGN/C\\
ALH09753&12:37:34.3&+62:20:56&2.12&-&104.77$\pm$10.27&0.934$\pm$0.151&24.133$\pm$0.208&42.50&-&-&-&1.85&1.7&4.50$\pm$0.06&-&AGN/C\\
ALH09821&12:37:42.4&+62:21:02&2.26&-&92.53$\pm$9.37&0.846$\pm$0.168&24.393$\pm$0.258&42.38&-&-&-&1.66&7.6&5.03$\pm$0.06&-&SF\\
ALH10766&12:37:36.4&+62:22:01&2.28&-&251.40$\pm$21.39&1.649$\pm$0.325&24.949$\pm$0.331&42.76&-&-&-&1.49&1.2&-&-&SF\\
ALH11687&12:37:18.4&+62:23:00&2.31&-&115.63$\pm$11.07&1.007$\pm$0.132&24.266$\pm$0.247&42.60&0.256$\pm$0.004&-&-&0.89&1.8&3.81$\pm$0.09&670.84$\pm$62.20&AGN/C\\
\hline
\end{tabular*}
\tablefoot{\\
\tablefoottext{1}{ID from ALHAMBRA survey data}\\
\tablefoottext{2}{Photometric redshift}\\
\tablefoottext{3}{Spectroscopic redshift from Barger et al. \cite{barger08}}\\
\tablefoottext{4}{Rest-frame equivalent width (\AA) of Ly$\alpha$ emission line from OFF-ON color and relation in Fig. \ref{filter}}\\
\tablefoottext{5}{OFF-ON color}\\
\tablefoottext{6}{ON-band magnitude}\\
\tablefoottext{7}{Logarithm of Ly$\alpha$ luminosity (erg s$^{-1}$)}\\
\tablefoottext{8}{Wavelength flux density in 24 $\mu{\rm m}$ band from MIPS ($\times$10$^{-18}$ erg cm$^{-2}$ s$^{-1}$ \AA$^{-1}$)}\\
\tablefoottext{9}{Wavelength flux density in 100 $\mu{\rm m}$ band from HERSCHEL-PACS ($\times$10$^{-18}$ erg cm$^{-2}$ s$^{-1}$ \AA$^{-1}$)}\\
\tablefoottext{10}{Wavelength flux density in 160 $\mu{\rm m}$ band from HERSCHEL-PACS ($\times$10$^{-18}$ erg cm$^{-2}$ s$^{-1}$ \AA$^{-1}$)}\\
\tablefoottext{11}{Isophotal (2-$\sigma$) radius in kpc}\\
\tablefoottext{12}{S\'ersic index}\\
\tablefoottext{13}{Observed R-Ks color}\\
\tablefoottext{14}{Ratio of $24\ \mu{\rm m}$ to R-band frequency flux densities}\\
\tablefoottext{15}{Spectral type from templates of Polletta et al. \cite{polletta07}: Star-forming (SF) or AGN/Composite (AGN/C)}\\
}
\end{table}
\end{landscape}

\begin{table*}[!h]
\caption{Number counts of LAEs redshifts 2.2 and 3.1, limited in Ly$\alpha$ luminosity and rest-frame EW. Note that only one AGN
out of the six observed in this work is detected in X-rays (see main text).}
\label{AGNfractionTable}
\setlength{\tabcolsep}{0.1cm}
\begin{tabular*}{9.6cm}{cccc}
\hline\hline
Reference&Mean redshift&Total number&AGN-to-total fraction\\
\hline
Gronwall et al. 2007  &3.1&64&0.016 \tablefootmark{1}  \\
Ouchi et al. 2008  &3.1&40&0.075\tablefootmark{1} \\
Nilsson et al. 2009  &2.25&73&0.123 \tablefootmark{1} \\
This work  &2.2&16&0.75 \tablefootmark{2} \\
\hline
\end{tabular*}
\tablefoot{\\
\tablefoottext{1}{X-ray selected sample}\\
\tablefoottext{2}{SED selected sample}\\
}
\end{table*}

\begin{figure*}[!h]
\centering
\resizebox{\hsize}{!}{\includegraphics{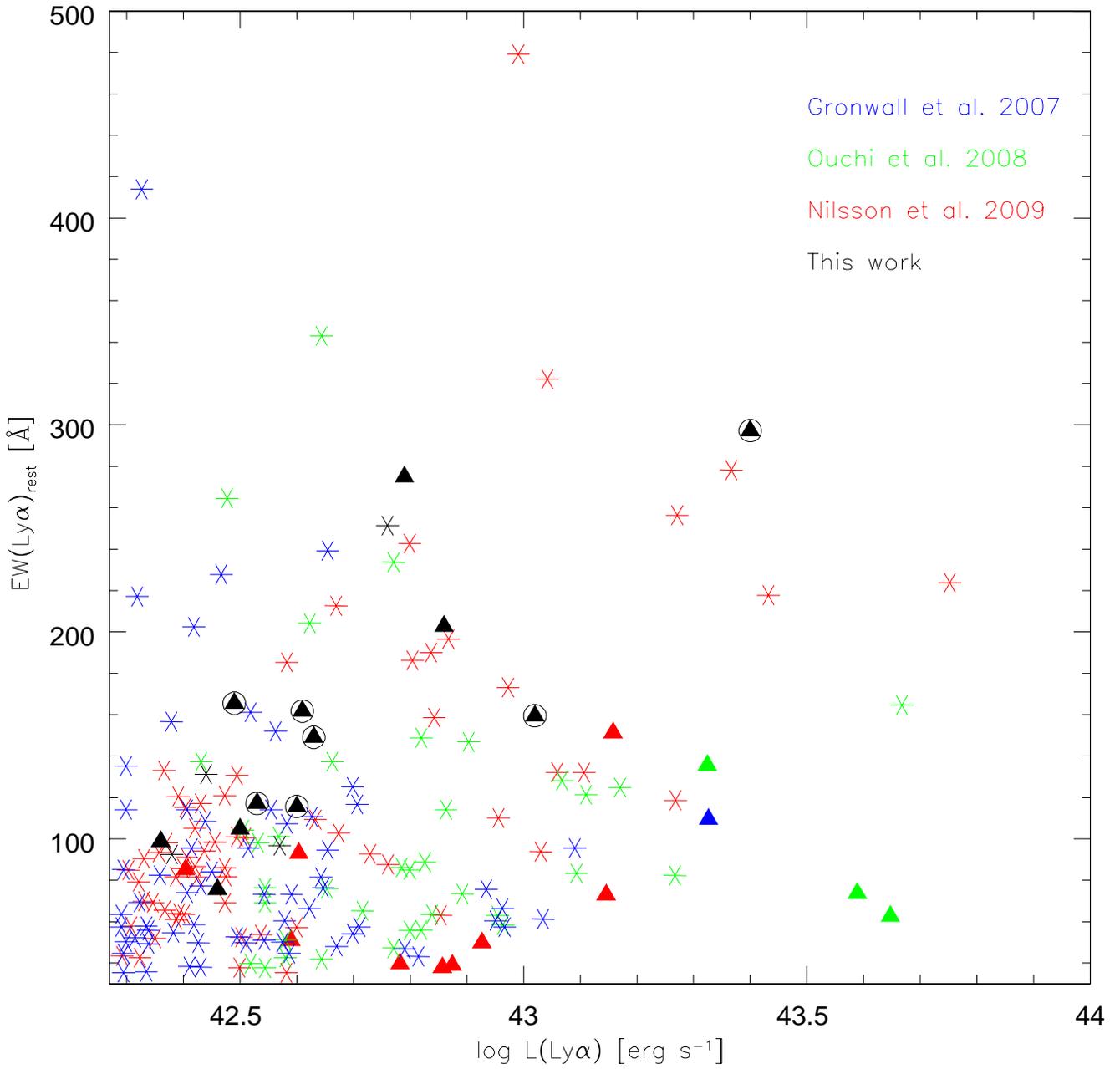}}
\caption{Rest-frame EW against Ly$\alpha$ luminosity for the LAE samples referred in Table \ref{AGNfractionTable}.
Star-forming like LAEs are represented by stars and the AGNs (or AGN/Composite objects) by triangles. Circled triangles 
are LAEs with FIR counterparts from this work. An interpretation of the AGN distribution in this diagram is given in 
the main text.} 
\label{AGNfractionPlot}
\end{figure*}

\end{document}